# Role of Carbon in Enhancing the Performance of $MgB_2$ superconductor

**V.P.S. Awana[1,2,*], Arpita Vajpayee[1], Monika Mudgel[1], Rajeev Rawat[3], Somobrata Acharya[2], H. Kishan[1], E. Takayama-Muromachi[4], A. V. Narlikar[3] and I. Felner[5]**

[1] National Physical Laboratory, Dr K.S. Krishnan Road, New Delhi-110012, India

[2] ICYS centre National Institute for Material Science (NIMS), Tsukuba, Ibaraki 305-0044, Japan

[3]UGC-DAE Consortium for Scientific Research, University Campus, Khandwa Road, Indore-452017, MP, India

[4]Advanced nano-materials Laboratory, National Institute for Material Science (NIMS), Tsukuba, Ibaraki 305-0044, Japan

[5]Racah Institute of Physics, the Hebrew University, Jerusalem, 91904, Israel

**Abstract**

The enhancement of the critical current density ($J_c(H)$) of carbon and *nano*-SiC doped $MgB_2$ is presented and compared. The upper critical field ($Hc_2$) being determined from resistivity under magnetic field experiments is though improved for both C substitution and *nano*-SiC addition the same is more pronounced for the former. In $MgB_{2-x}C_x$ carbon is substituted for boron that induces disorder in the boron network and acts as internal pinning centres. The optimal $J_c(H)$ values are obtained for x = 0.1 sample . In case of *nano*-SiC doped in $MgB_2$, the $J_c(H)$ improves more profoundly and two simultaneous mechanisms seems responsible to this enhancement. Highly reactive *nano*-SiC releases free carbon atom, which gets easily incorporated into the $MgB_2$ lattice to act as *intrinsic* pinning centres. Further enhancement is observed for higher *nano*-SiC concentrations, where the un-reacted components serve as additional *extrinsic* pinning centres.

* Corresponding Author: e-mail: awana@mail.nplindia.ernet.in

## 1. Introduction

Recently, there have been several reports describing high performance *nano* (n) - particle doped $MgB_2$ superconductor where the *n*-SiC/*n*-diamond/*n*-carbon-tubes/*n*-carbon have yielded high dividends [1-5]. One wonders if the enhanced flux pinning is indeed solely due to these small size (< 10 *nm*) additives or whether carbon plays any additional important role.

Soon after the discovery of superconductivity in $MgB_2$ at $T_c$ nearly 40 K [6], its various interesting physical properties were found to be intimately linked with: (a) the unexpected transparency of its grain-boundaries [7,8] to passage of electric current, and (b) to the existence of unusual two-band structure [9]. Because of its higher $T_c$, $MgB_2$ clearly holds a distinct advantage over most other superconductors in that it can conveniently function at a relatively higher temperature of, say 20K, easily produced using a conventional close cycle cooling system. Although the $T_c$ of high temperature superconductors(*HTS*) is much higher, i.e., around 90-130 K, their small coherence length $\xi$ (around 0.3 nm to 2.0 nm), large anisotropy and ceramic nature have posed formidable problems for conductor development for practical applications. It is nearly impossible to optimally pin the flux vortices in *HTS* via foreign additives because of very small $\xi$. On the other hand, the relatively larger $\xi$ of $MgB_2$ (about 5-10 nm) allows optimum pinning by *n*-particles for enhancement of the $J_c(H)$ performance. Although this effect has been studied for a variety of different *nano*-particle additives, e.g., *n*-$Dy_2O_3$ [10], *n*-$SiO_2$ [11], *n*-carbon [2,4], *n*-tubes of carbon [3] and *n*-diamond [5], the best performance yet seen, is with nanosized carbon derivatives. In fact it might be interesting to inter-compare the high field critical current ($J_c$,$H$) behaviour resulting from carbon, when substituted for boron ($MgB_{2-x}C_x$) with that when it (or its derivatives) are introduced as additive. We found here, that in the $MgB_{2-x}C_x$ system, though $T_c$ decreases with x, the high field $J_c(H)$, when measured at lower temperatures, is markedly higher for low x values. For $MgB_2$+ *n*-SiC the $T_c$ decreased slightly but the $J_c(H)$ goes on improving more profoundly till 7 %wt addition. We thus conclude that the enhancement in $J_c(H)$ is a result of two parameters: the existence of pinning centres due to addition of *n*-SiC and, simultaneously, the presence of disorder in boron networks caused by carbon substitution. Our present results demonstrate the importance of disorder originated from carbon substitution.

## 2. Experimental details

Our polycrystalline $MgB_2$-*n*SiC (n= 0, 5%, 7% & 10%) and $MgB_{2-x}C_x$ (x= 0, 0.04, 0.08 0.10 & 0.20) samples were synthesized by solid-state reaction route with ingredients of Mg, B, *n*-SiC and C powders. The Mg powder used is from *Reidel-de-Haen* of 99% purity, insoluble in HCl and with Fe impurity of less than 0.05%. Amorphous B powder used is from *Fluka* (of 95-97% purity). The *n*-SiC powder is from Aldrich with average particle size (*APS*) of 5-12 nm. For synthesising the samples, the stoichiometric amounts were ground thoroughly, palletized, and put in a quartz tube inserted at 850 $^{0}$C (heating rate 425 $^{0}$C per hour) under a flow of argon at ambient pressure. This temperature was held for 2.5 hours, and subsequently cooled under Ar, to room temperature over a timeframe of 6 hours. The x-ray diffraction pattern of the compound was recorded by using $CuK_{\alpha}$ radiation. *Jeol*-200KV transmission electron microscope was used for bright/dark field images. The magnetization measurements were carried out using *Quantum Design MPMS-XL* magnetometer.

## 3. Results and discussion

Fig. 1 depicts the X-ray diffraction (*XRD*) patterns of some of the studied $MgB_{2-x}C_x$ and *n*-SiC added $MgB_2$ samples and the lattice parameters are listed in Table 1.The pristine $MgB_2$ compound carries a small amount of un-reacted MgO (marked as #). In $MgB_{2-x}C_x$ the single phase is retained up to x = 0.2. When *n*-SiC is added to $MgB_2$, the phase purity is maintained up 3-wt%. For 5-wt% (assigned as *n*-SiC5) the SiC derivative viz. $Mg_2Si$ lines are clearly seen in the *XRD* spectrum (marked by *). Lattice parameters are given in Table 1. For $MgB_{2-x}C_x$ system thought the *a*-lattice parameter decreases, the *c* increases slightly with increasing x up to x = 0.2. The same is true for *n*-SiC added samples except that some un-reacted lines are seen above 5-wt% addition. This is consistent with previous reported data on $MgB_{2-x}C_x$ [12]. This indicates clearly that in both cases carbon enters to the matrix. The nano *n*-SiC reacts with Mg (leaving free silicon or $Mg_2Si$) and release highly reactive free C which can be easily incorporated into the lattice of $MgB_2$ and substitute into the B sites [13]. For higher *n*-SiC concentrations, the unbroken *n*-SiC still acts as additional pinning centres. The inset in Fig. 1 shows the presence of *nano* particles in bright field *TEM* image with average particle size of around 10 *nm*.

The zero-field-cooled (ZFC) and field-cooled (FC) magnetic susceptibility curves of the $MgB_{2-x}C_x$ and *n*-SiC added $MgB_2$ samples are shown in Fig. 2. The $T_c$ values for all the studied samples are given in Table 1. The ZFC branch of pristine $MgB_2$ exhibits a sharp diamagnetic superconducting transition ($T_c$) at 38.5 K, whereas the FC signal is very small due to intrinsic pinning. Intrinsic pinning in $MgB_2$ due to various *nano*-structural defects was highlighted by us very recently [14] and is not discussed here. For $MgB_{2-x}C_x$ system, the $T_c$ values of 34.5 and 32 K for with x = 0.10 and 0.20 are obtained, which are in agreement with the previous reports [12]. For *n*-SiC added $MgB_2$ materials the $T_c$ obtained for 5-wt % is 36 K and 35 K for 10wt% *n*-SiC added sample. The relatively constant $T_c$ in *n*-SiC added $MgB_2$ samples for 5 wt% up to 10 wt% indicates that at higher concentrations, only a fraction of C presumably substitutes for boron in $MgB_2$, which would justify the marginal decrease in $T_c$. The rest of *n*-SiC is available in the matrix and serves as pinning centres.

Fig. 3 depicts the $J_c(H)$ results for *n*-SiC added $MgB_2$ samples at 10 and 20 K up to relatively high fields of 70 kOe. The critical current is deduced from the $M(H)$ plots by invoking Bean's critical state model. As inferred from this figure and Table 1, $J_c(H)$ performance of $MgB_2$+7 wt% *n*-SiC sample is the optimum one at both temperatures and its value at 10 K under 70 kOe field is an order of magnitude higher than the pristine $MgB_2$. To see the irreversibility field ($H_{irr}$), the extended $M(H)$ plots at 10 K of these samples are shown in the inset of Fig. 3. As may be seen, for pristine $MgB_2$, the $H_{irr}$ is close to 70 kOe and it is significantly enhanced for *n*-SiC added samples. Note the higher $H_{irr}$ for the 7 wt% *n*-SiC added sample and the relatively smaller loop for the 10 wt% sample, which indicates that the 7 wt% *n*-SiC is optimum concentration $MgB_2$ performance. The *"curl"* like curve exhibited in Figs. 3 and 4, are due to flux jumps which are observed at low applied fields, only for higher $J_c(H)$ samples and very low heat capacity. The flux jumps phenomenon was already commented very recently by some of us in an earlier report [15].

The $J_c(H)$ results for $MgB_{2-x}C_x$ samples along with pristine $MgB_2$ are shown in Fig. 4. The low field region (< 40 KOe), $J_c(H)$ of pristine $MgB_2$ is higher than that of $MgB_{2-x}C_x$: x = 0.10 and 0.20 samples. However as field increases further, the $J_c(H)$ performance of C doped samples is better than of that of pure $MgB_2$. The sample with x = 0.10 (i.e. the 5at% C at B site in $MgB_2$) has the best characteristics. Its $J_c$ at 10 K and 70 kOe, is higher than for both x = 0 and x = 0.20 samples, in spite of the reduction in $T_c$ by 4 K (Table 1). Comparatively, the $J_c(H)$ performance of $MgB_2$ + *n*-SiC (Fig.3) is superior than that of $MgB_{2-x}C_x$ (Fig.4).

To confirm the improved flux pinning behaviour through carbon doping, the field dependence of normalized flux pinning force ($F_p / F_{p,\,max}$) is shown in fig. 5 & fig. 6 at 20 K. The relationship between flux pinning force and critical current density could be described by [16, 17].

$$Fp = \mu_0 J_c(H)\, H \qquad (1)$$

Where $\mu_0$ is the magnetic permeability in vacuum. It is clear from these figures that the pinning forces are much higher than the pure $MgB_2$ above 1.5 Tesla for both C and SiC doped samples, indicating enhanced flux pinning force in high fields. It can be seen in fig. 5 that for the 7-wt% *n*-SiC doped sample the peak is much broader than those of other three samples, indicating that the pinning strength in the higher field region has improved. The same behaviour is shown by $MgB_{2-x}C_x$: x = 0.10 sample. The Broadness of the $F_p / F_{p,\,max}$ versus $H$ plot is comparatively more for $MgB_2$ + *n*-SiC (Fig.5) than that of C doped (Fig.6) samples. This indicates that higher field flux pinning is more effective in *n*-SiC doped samples than in $MgB_{2-x}C_x$.

Resistance variation with temperature of Pure, SiC 7-wt% and C (x=0.10 in $MgB_{2-x}C_x$) added $MgB_2$ samples are shown in fig.7. From this figure we see that the room temperature resistance value is increased as we doped C & SiC in $MgB_2$. The residual resistivity ratio (RRR = $R_{T280K}/R_{Tonset}$) values for these samples are obtained as 3.15, 1.73 & 1.53 respectively. It indicates towards the increased disorder in doped samples in comparison to pure one. Transition temperature ($T_c$) decreases in both SiC added and C substituted sample as compared to pure $MgB_2$ because of carbon substitution at boron site, which is confirmed by decrease in lattice parameter '*a*'. Exact values of $T_c(R=0)$ and lattice parameters are tabulated in Table 1.

Transition zone of Resistance vs Temperature plots at varying field values is shown for Pure, SiC 7% and C (x=0.10 in $MgB_{2-x}C_x$) added $MgB_2$ samples in Fig. 8. Transition width is more or less same in all the samples. With the help of this graph we found out the upper critical field values and the variation of upper critical field with temperature for SiC 7%, C (x=0.08 & 0.10 in $MgB_{2-x}C_x$) and pure $MgB_2$ sample is shown in fig. 9. This figure depicts that there is an increment in $H_{c2}$ values by both the dopants because of carbon substitution [18]. Increase in $H_{c2}$ is accompanied by a rise in resistivity. Relative increase in $H_{c2}$ is more for $MgB_{2-x}C_x$ than that of pure and *n*-SiC doped samples at all temperature. It reveals that carbon substitutes at Boron site in $MgB_2$ lattice more prominently in case of $Mg_{1-x}B_{2-x}C_x$ as compared to *n*-SiC added samples. This is indicated by steep decrease in *a* parameter (*XRD* results) and *Tc* of $Mg_{1-x}B_{2-x}C_x$ samples. On the other side in $MgB_2$ + *n*-SiC series of samples, some free Carbon being available from broken *n*-SiC also substitutes at Boron site but in less effective manner than in $Mg_{1-x}B_{2-x}C_x$, resulting in a relatively smaller decrease in *a* parameter and $T_c$. It seems that only a portion of broken *n*-SiC goes to the grain boundary and/or within the $MgB_2$ grains as *nano*-inclusions and remaining unbroken *n*-SiC and $Mg_2Si$ particles act as extrinsic pinning centres in the composite system. This results in better *Jc*(*H*) performance for $MgB_2$ + *n*-SiC samples than that of $Mg_{1-x}B_{2-x}C_x$. Summarily though the upper critical field values are higher for $Mg_{1-x}B_{2-x}C_x$, the Jc(H) performance is better for $MgB_2$ + *n*-SiC series of samples. This is primarily due to the dual role of added *n*-SiC, first as an extrinsic *nano*-pinning agent and secondly its broken free Carbon which substitutes at B-site in $MgB_2$ lattice and acts separately as intrinsic pinning disorder in the system. Both these effects improve the performance more dramatically than as for $Mg_{1-x}B_{2-x}C_x$. Lattice parameters, critical temperatures ($T_c$) and critical current density ($J_c$) values are tabulated in Table 1.

For quantitative comparison, the $J_c(H)$ (50 KOe, 10 K) values for all the studied $MgB_{2-x}C_x$ and *n*-SiC added $MgB_2$ samples are given in Table 1. It appears that the $J_c(H)$ values obtained for $MgB_2$ + *n*-SiC (Fig. 3), are much higher than that of the substituted $MgB_{2-x}C_x$ samples. In the later case, the enhancement of $J_c(H)$ is entirely due to the disorder in σ band [9,12] via carbon substitution at boron site, as there are no additives serving as *nano*-pins. On the other hand, in the $MgB_2$ + *n*-SiC system the higher $J_c(H)$ arises from additives which serve as pinning centres. So in principle the $J_c(H)$ of $MgB_2$ can be enhanced by two mechanisms: (1) by partial substitution (of about 5%) of boron by carbon as occurs in $MgB_{2-x}C_x$ and (2) by un-reacted nano-particles which serve as pinning centres. In this way we may understand why *n*-SiC, *n*-Diamond, *n*-Carbon-tubes and *n*-Carbon additions to $MgB_2$ yielded higher dividends than other additives of non-carbon *nano*-derivatives [1-5, 13, 19, 20]. In $MgB_{2-x}C_x$ as well as in low concentrations of *n*-SiC added materials the C substitution for B induces disorder in lattice sites and thereby contributes to additional flux pinning and substantiates and hence enhances the $MgB_2$ performance. Further increase of $J_c(H)$ occurs for high concentrations of *n*-SiC where the remaining unbroken *nano*-particles though in lesser concentration are still present within the $MgB_2$ grains.

Worth mentioning is the fact that breaking of doped *nano*-Carbon derivatives and ensuing substitution at B site in $MgB_2$ depends upon both nature of dopant and the heat treatment [13, 19-21]. For example, the *n*-SiC is more susceptible to react than *n*-Diamond [13, 19-21] at common $MgB_2$ synthesis temperatures.

## 4. Summary and Conclusion

To sum up, we have provided clear evidence that two mechanisms are responsible for the enhancement of $J_c(H)$ in carbon doped $MgB_2$. A nontrivial contribution to flux pinning originates from the disorder created in σ band superconducting condensate, particularly in the boron networks, serving as weakly superconducting pins. Beside that, un-reacted *nano*-particles introduced as additives serve as normal pin centres. Both components are responsible for the enhancement in $Jc(H)$ of $MgB_2$ formed with *nano*-carbon derivatives as additives.

**Acknowledgement**

This research is partly supported by the Israeli Ministry of Science under the mutual India (*DST*) - Israel (*MST*) grant. One of us (VPSA) thanks ICYS centre Director Prof. H. Bando for his current visit to NIMS Japan. AVN would like to thank Indian National Science Academy, New Delhi for the Senior Scientist position. Arpita Vajpayee and Monika Mudgel would like to thank the *CSIR* for the award of *JRF* to pursue their *Ph. D* degree. Authors from the *NPL* thank their director Professor Vikram Kumar for encouragement.

Table 1: The *a*, *c* lattice parameters, $T_c$ and $J_c$ (50 KOe, 10K) for studied $MgB_2$-*n*SiC and $MgB_{2-x}C_x$ samples

| **Sample** | ***a*** **(Å)** | ***c*** **(Å)** | $T_c^{dia}$ **(K)** | ***Tc(R=0)*** | $J_c$ **(A/cm²)** **(50 KOe/10K)** |
|---|---|---|---|---|---|
| $MgB_2$ | 3.0817(8) | 3.5230(8) | 38.5 | 38.2 | $5.40 \times 10^3$ |
| $MgB_{1.96}C_{0.04}$ | 3.0793(7) | 3.5280(8) | 37 | - | $1.27 \times 10^4$ |
| $MgB_{1.92}C_{0.08}$ | 3.0754(16) | 3.5275(16) | 36 | 34 | $1.10 \times 10^4$ |
| $MgB_{1.90}C_{0.10}$ | 3.0742(24) | 3.5287(24) | 34.5 | 32.5 | $1.30 \times 10^4$ |
| $MgB_{1.80}C_{0.20}$ | 3.0678(20) | 3.5336(21) | 32 | 29.5 | $7.3 \times 10^3$ |
| $MgB_2$+SiC5at% | 3.0762(5) | 3.5289(8) | 36 | 35.5 | $1.4 \times 10^4$ |
| $MgB_2$+SiC7at% | 3.0772(9) | 3.5297(6) | 36 | 35 | $4.15 \times 10^4$ |
| $MgB_2$+SiC10at% | 3.0759(8) | 3.5292(9) | 35 | 34.5 | $3.07 \times 10^4$ |

*Figure Captions*:

Fig. 1: X-ray diffraction patterns at room temperature of various $MgB_2$-*n*SiC and $MgB_{2-x}C_x$ samples. The MgO impurity is marked as # and Si derivatives with *. The inset shows the *TEM* image with scale bar of 100 nm, for $MgB_2$+7wt% *n*-SiC sample, highlighting the presence of *nano*-particles in the matrix.

Fig. 2: Low field (10 Oe) magnetic susceptibility ($\chi$) versus Temperature ($T$) plots for various $MgB_2$-*n*SiC and $MgB_{2-x}C_x$ samples.

Fig. 3: $J_c(H)$ plots at 10 and 20 K for $MgB_2$-*n*SiC samples, the inset shows the extended $M(H)$ plots of the same to mark the $H_{irr}$.

Fig. 4: $J_c(H)$ plots at 10 and 20 K for $MgB_{2-x}C_x$ samples; the notation C4, C8, C10 and C20, does mean x = 0.04, 0.08, 0.10 and 0.20 respectively.

Fig. 5: Reduced flux pinning force ($F_p/F_{p,max}$) variation with magnetic field for SiC doped samples at 20 K

Fig. 6: Reduced flux pinning force ($F_p/F_{p,max}$) variation with magnetic field for $MgB_{2-x}C_x$ samples at 20K; the notation C4, C8, C10 and C20, does mean x = 0.04, 0.08, 0.10 and 0.20 respectively.

Fig.7: Resistance vs Temp. Plots for $MgB_{2-x}C_x$, x=0.10, $MgB_2$+nSiC7% and Pure $MgB_2$ samples

Fig. 8: Transition zone of resistance vs Temperature Plots under different magnetic fields up to 8 Tesla for $MgB_{2-x}C_x$, x=0.10, $MgB_2$+nSiC7% and Pure $MgB_2$ samples

Fig. 9: Upper critical Field ($H_{c2}$) vs Normalized temperature Plots for $MgB_{2-x}C_x$, x=0.10 & x=0.08, $MgB_2$+nSiC7% and Pure $MgB_2$ samples

Fig. 1 Awana et al,

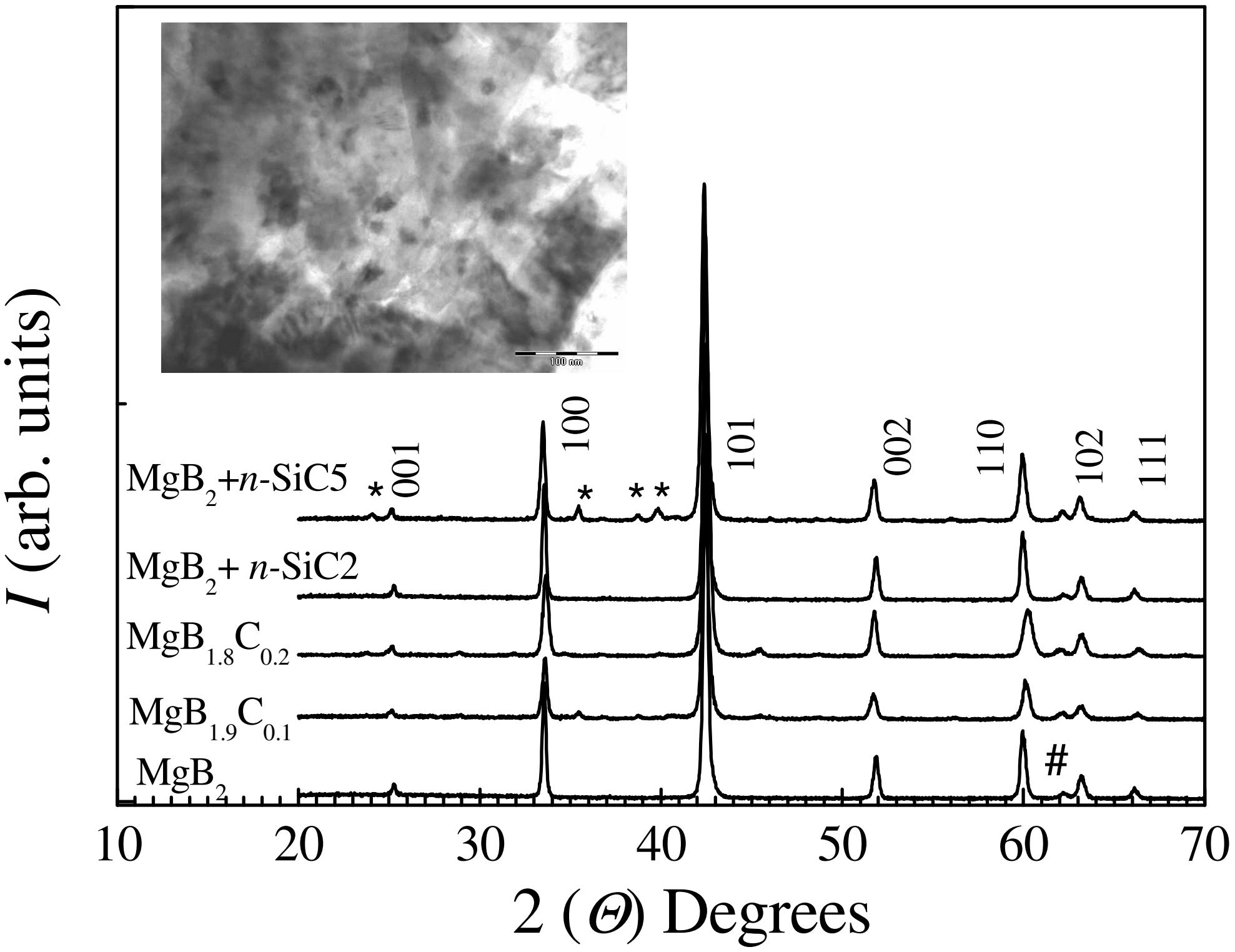

Fig. 2 Awana et al,

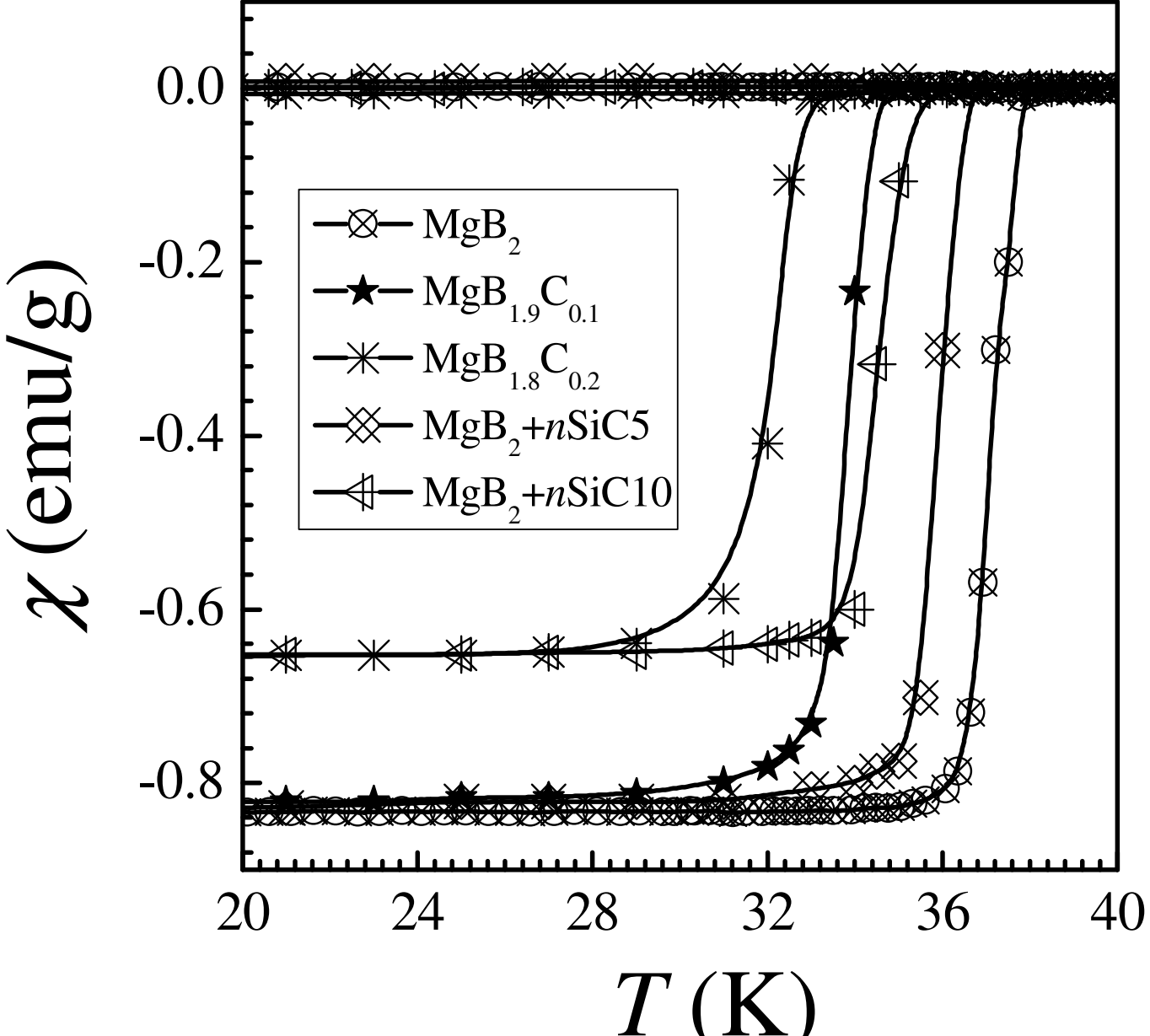

Fig. 3 Awana et al,

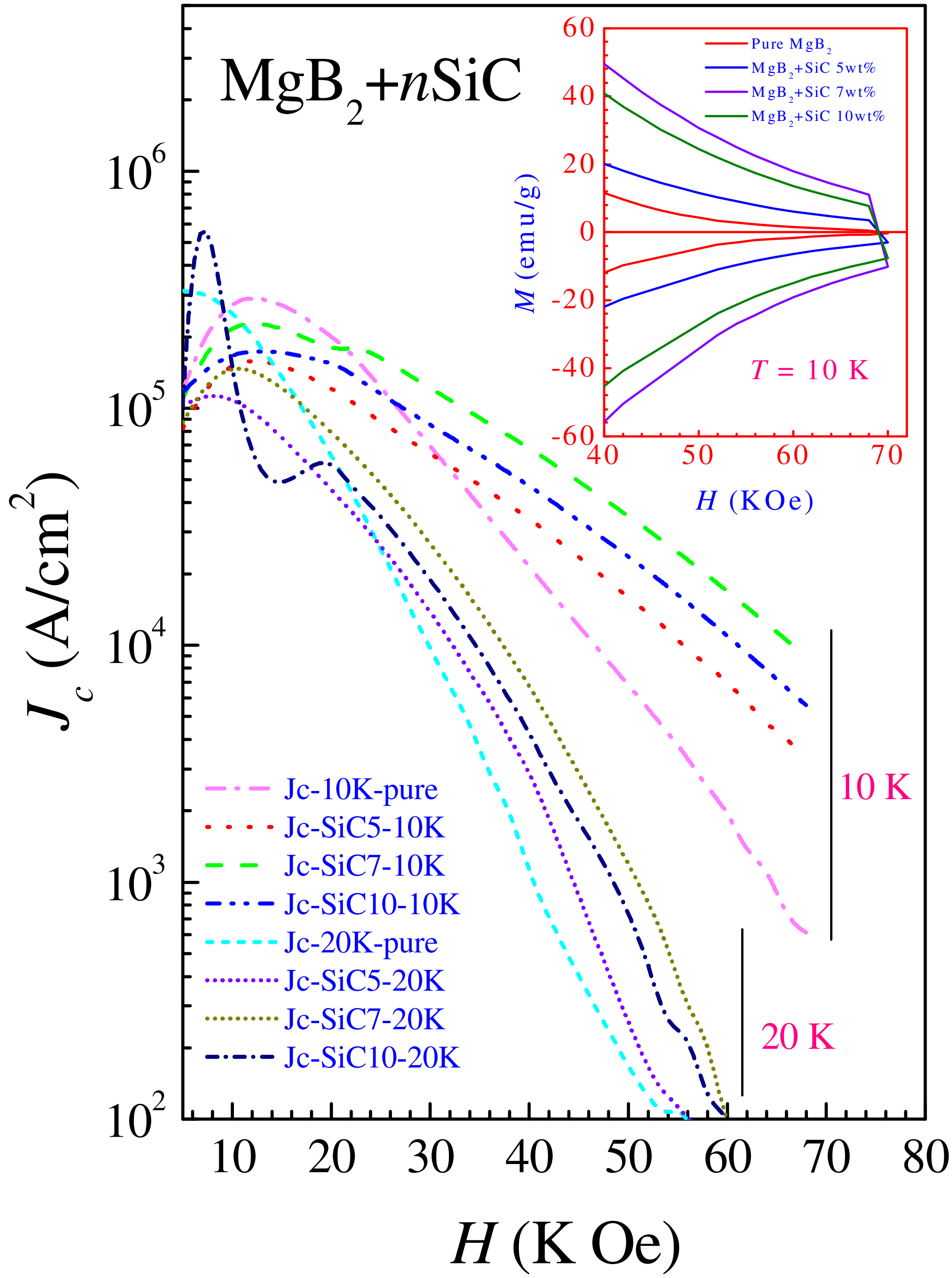

Fig. 4 Awana et al,

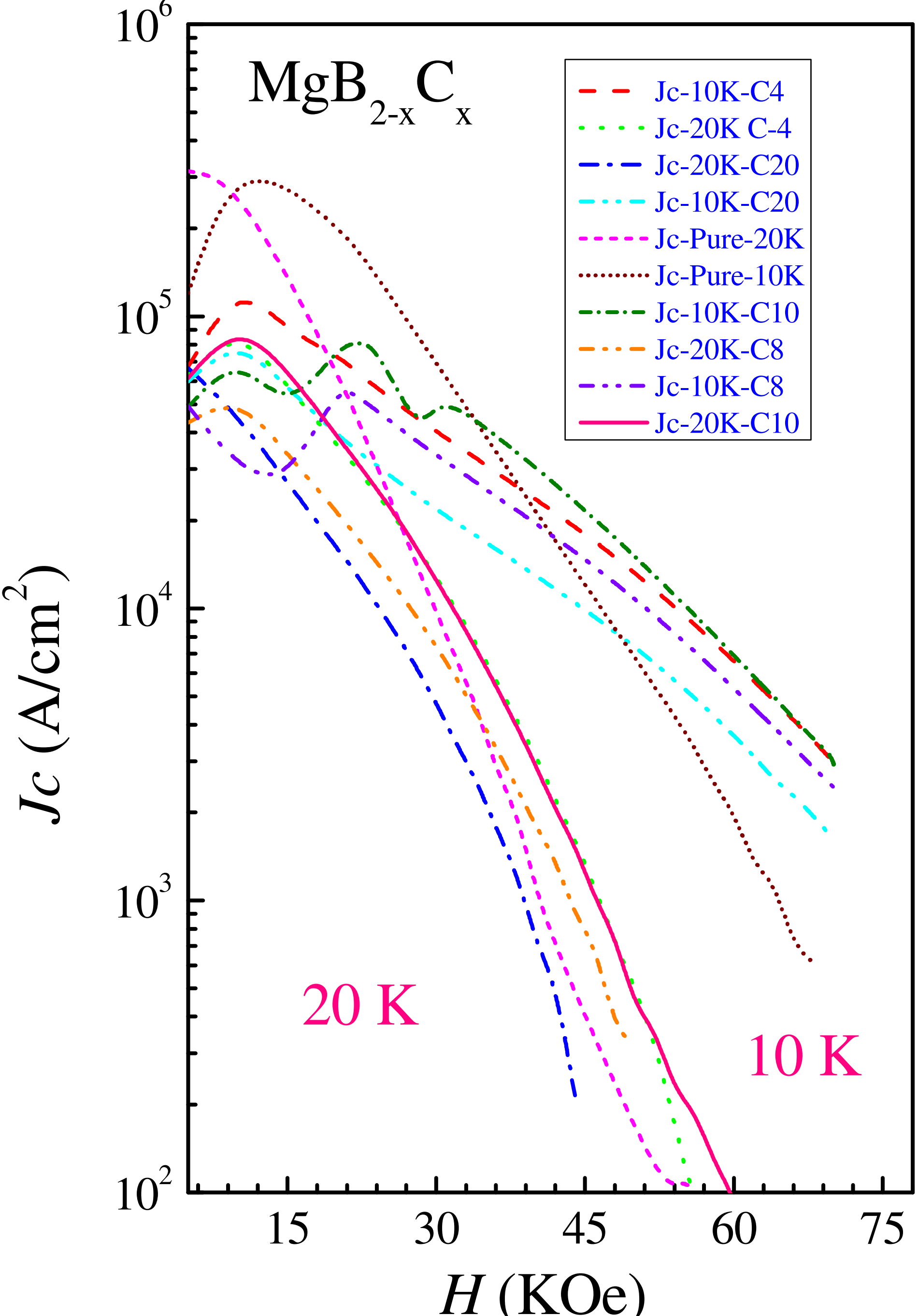

Fig. 5 Awana et al,

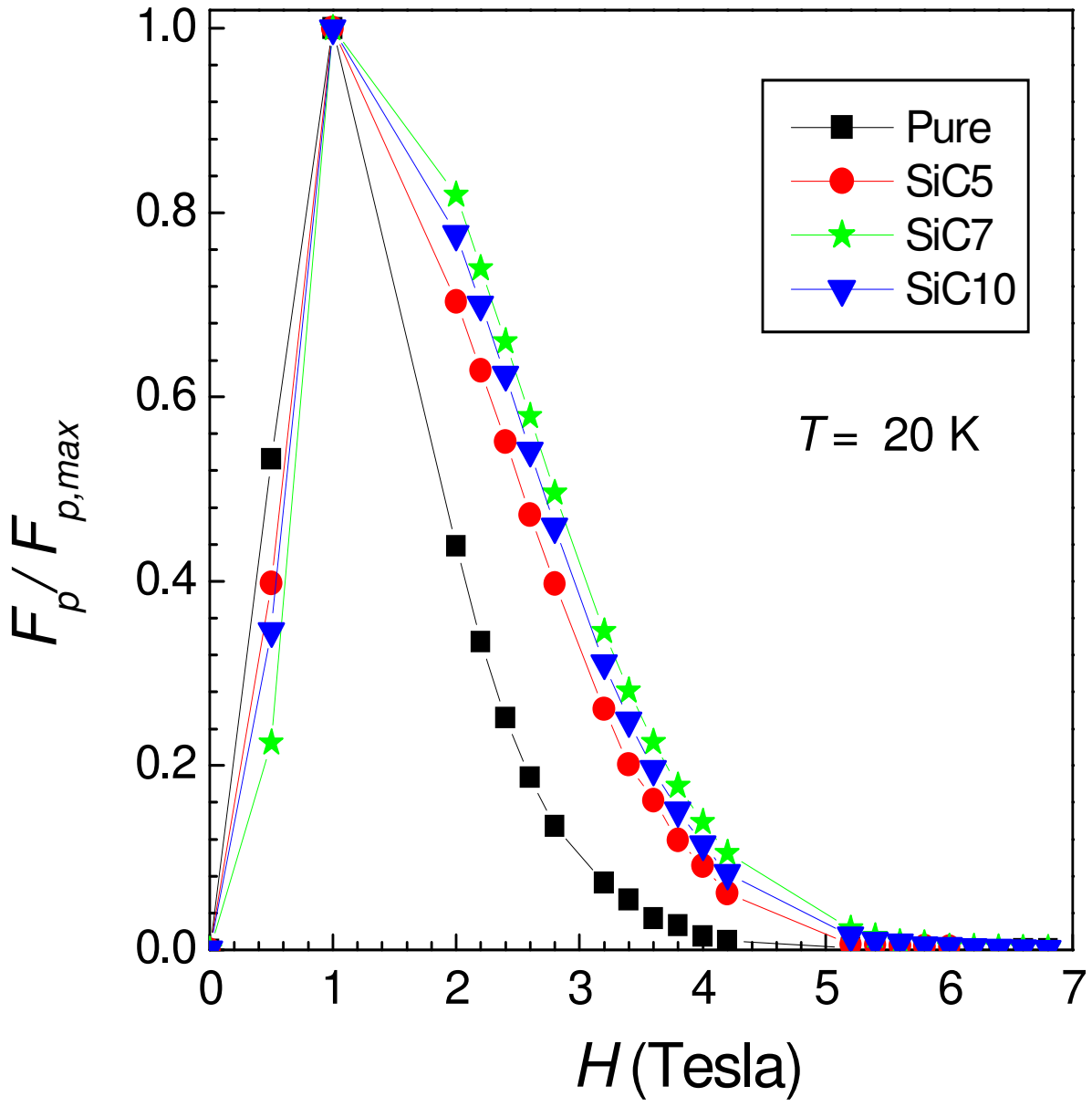

Fig. 6 Awana et al,

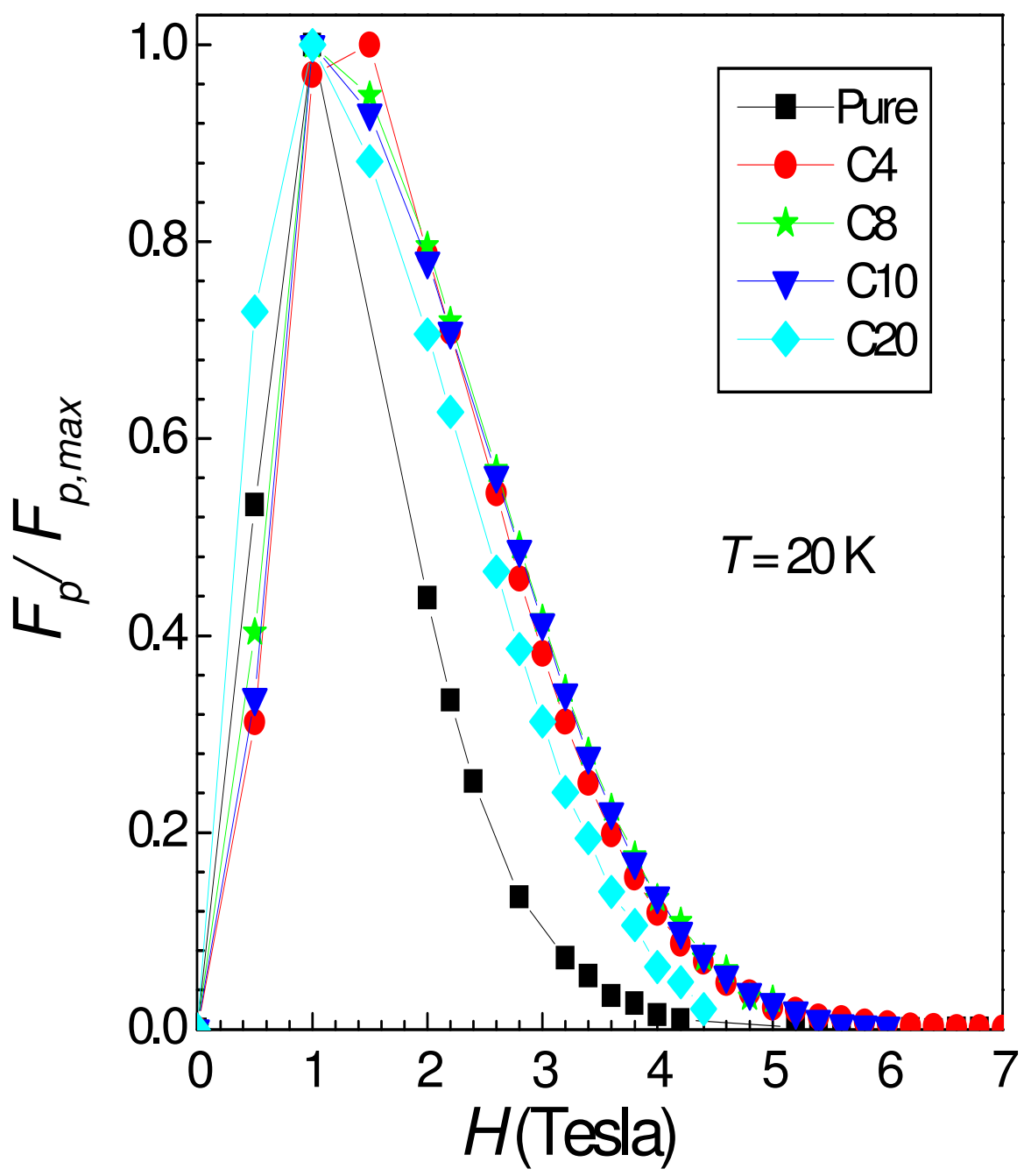

Fig. 7 Awana et al,

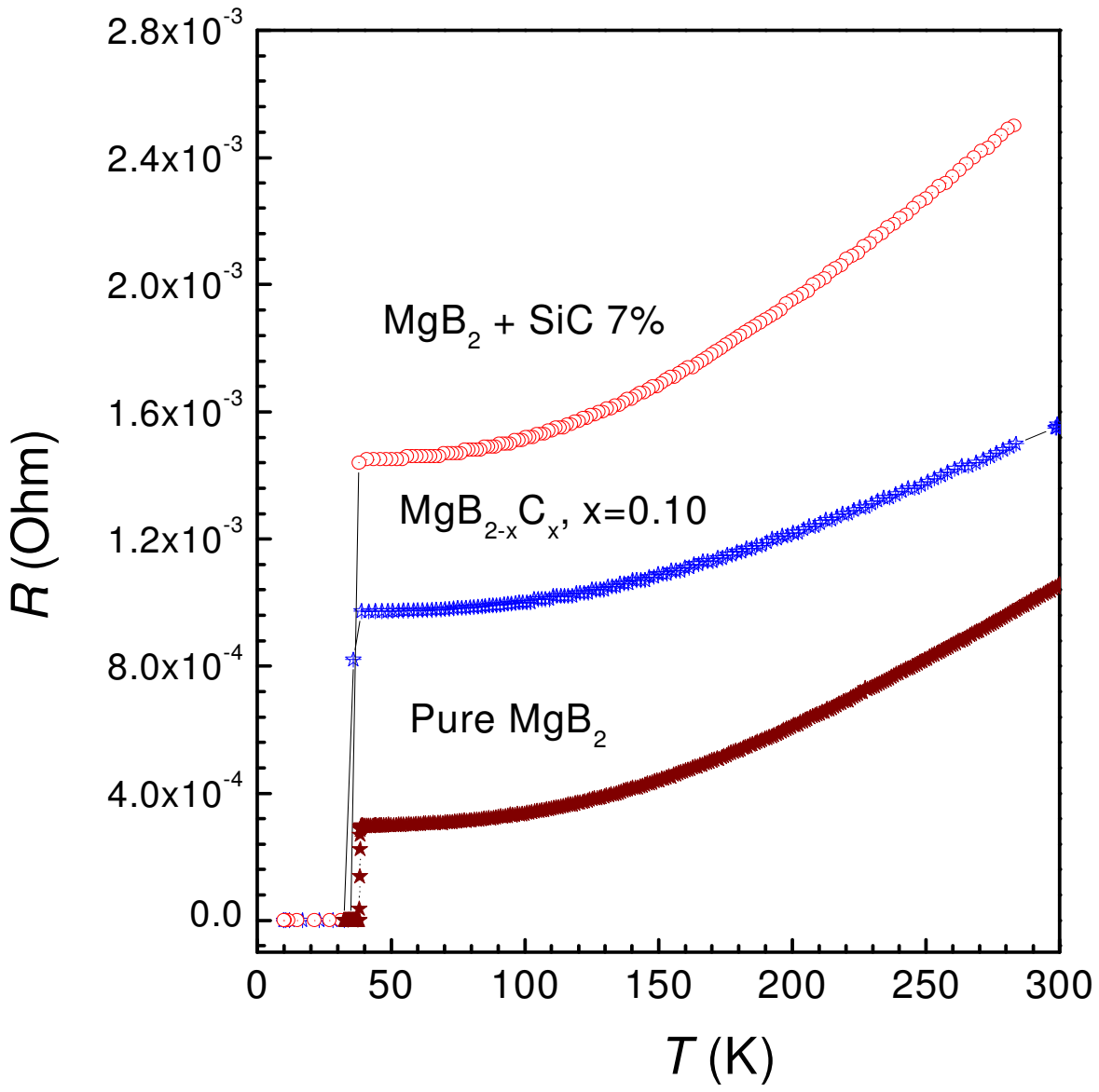

Fig. 8 Awana et al

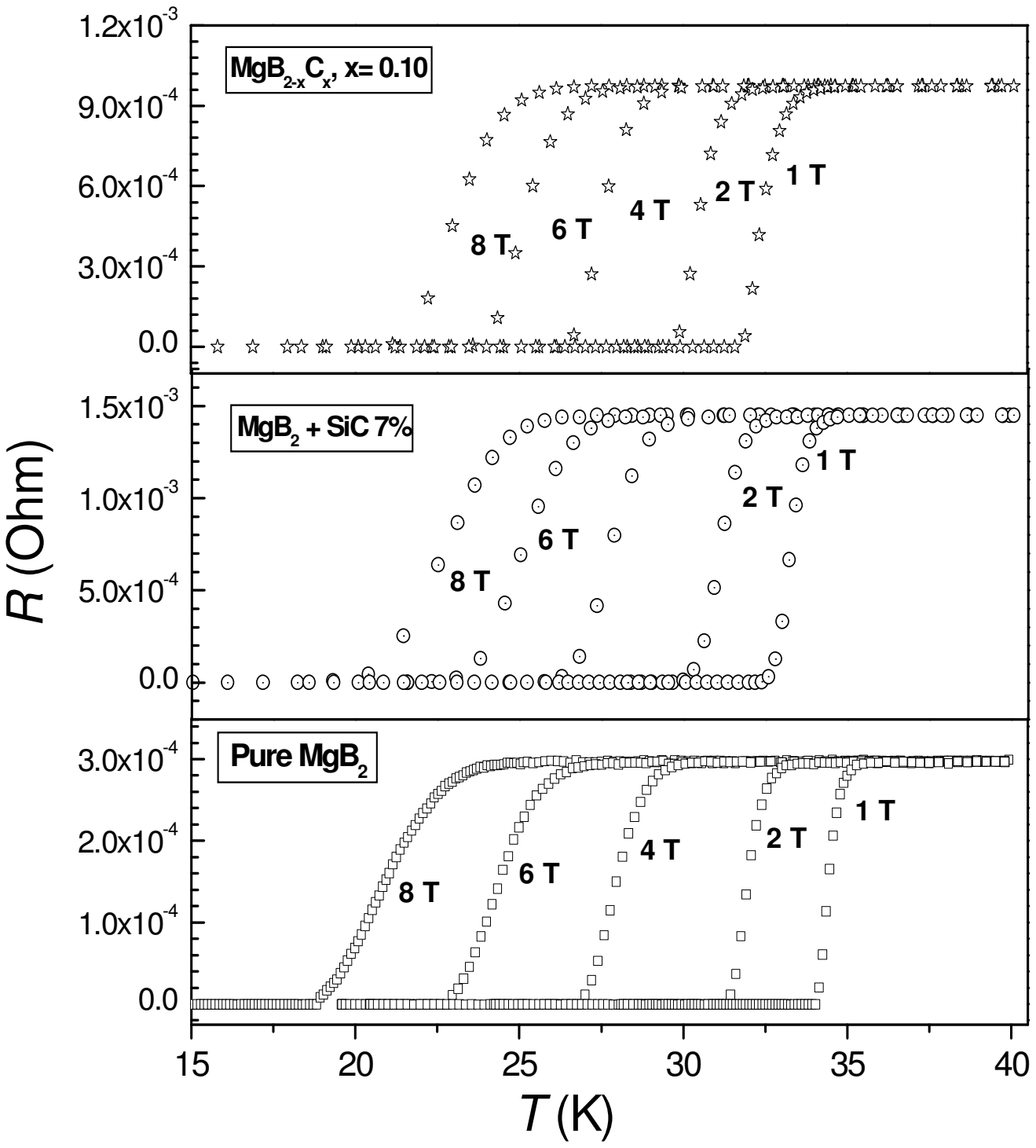

Fig. 9 Awana et al

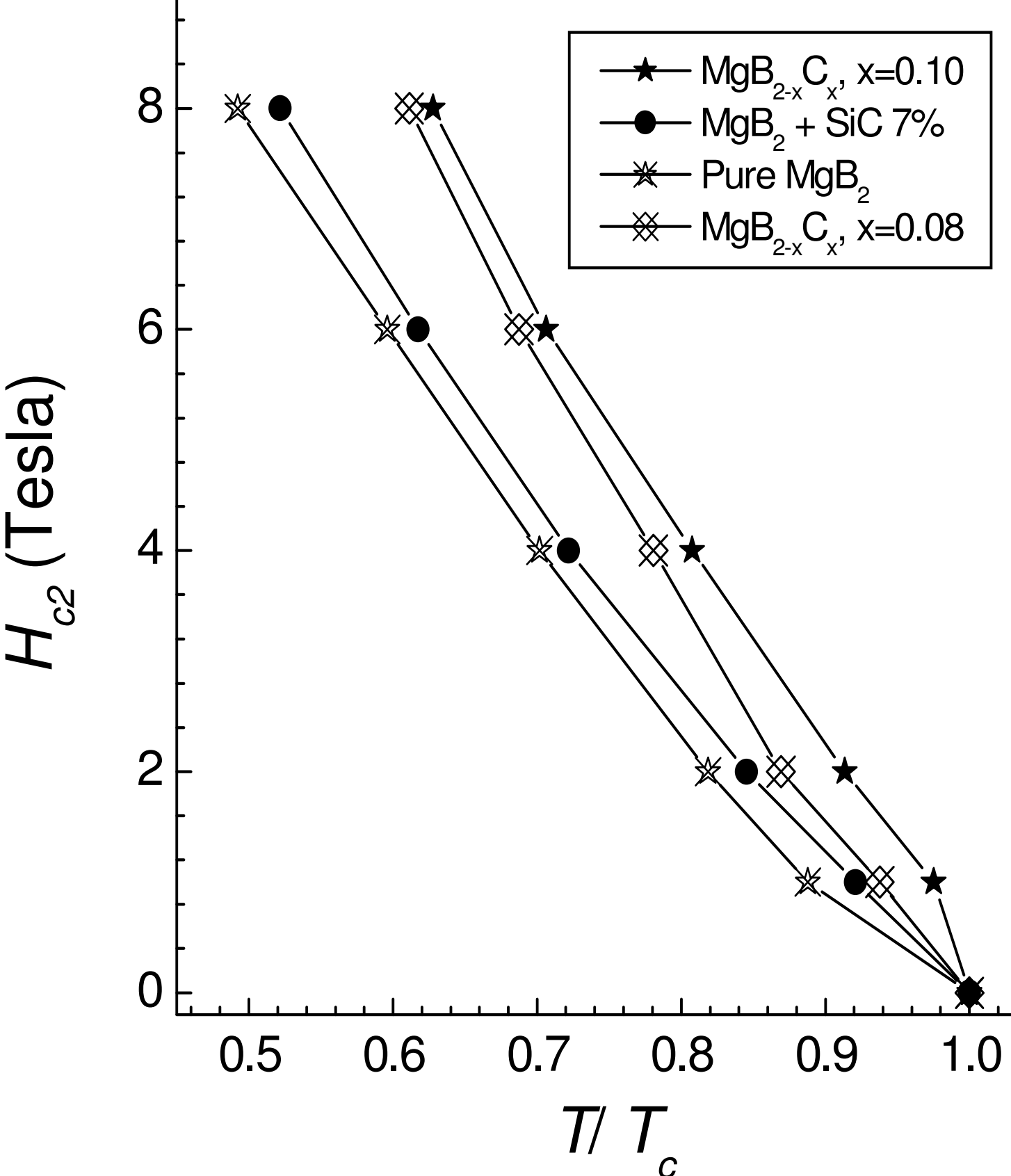